\documentclass[fleqn,usenatbib]{mnras}

\usepackage{newtxtext,newtxmath}
\usepackage[T1]{fontenc}
\usepackage{graphicx}
\usepackage{amsmath,scalerel}
\usepackage{relsize}
\usepackage{subcaption}
\usepackage{soul}
\usepackage{booktabs}
\usepackage{comment}
\usepackage{ulem}

\DeclareRobustCommand{\VAN}[3]{#2}
\let\VANthebibliography\thebibliography
\def\thebibliography{\DeclareRobustCommand{\VAN}[3]{##3}\VANthebibliography}

\definecolor{sk}{rgb}{0.99, 0.00, 0.5}

\title[Downsizing stops in the dwarf regime]{Downsizing does not extend to dwarf galaxies: identifying the stellar mass regimes shaped by supernova and AGN feedback}

\author[I. Lazar et al.]{I. Lazar\thanks{E-mail: i.lazar@herts.ac.uk},$^{1}$ S. Kaviraj,$^{1}$ G. Martin,$^{2}$ C. J. Conselice,$^{3}$ S. Koudmani,$^{1,4}$ A. E. Watkins,$^{1}$
S. K. Yi,$^{5}$ \newauthor D. Kakkad,$^{1}$, T. M. Sedgwick$^{1}$, Y. Dubois,$^{6}$ J. E. G. Devriendt,$^{7}$ K. Kraljic$^{8}$ and S. Peirani$^{9,10}$\\
$^{1}$Centre for Astrophysics Research, Department of Physics, Astronomy and Mathematics, University of Hertfordshire, College Lane, Hatfield AL10 9AB, UK\\
$^{2}$School of Physics and Astronomy, University of Nottingham, University Park, Nottingham NG7 2RD, UK\\
$^{3}$Jodrell Bank Centre for Astrophysics, University of Manchester, Oxford Road, Manchester M13 9PL, UK\\
$^{4}$St Catharine's College, University of Cambridge, Trumpington Street, Cambridge CB2 1RL, UK\\
$^{5}$Department of Astronomy and Yonsei University Observatory, Yonsei University, 50 Yonsei-ro, Seodaemun-gu, Seoul 03722, Republic of Korea\\
$^{6}$Institut d'Astrophysique de Paris, Sorbonne Universit\'es, UMPC Univ Paris 06 et CNRS, UMP 7095, 98 bis bd Arago, 75014 Paris, France\\
$^{7}$Department of Physics, University of Oxford, Keble Road, Oxford OX1 3RH UK\\
$^{8}$Observatoire Astronomique de Strasbourg, Universit\'e de Strasbourg, CNRS, UMR 7550, F-67000 Strasbourg, France\\
$^{9}$ILANCE, CNRS – University of Tokyo International Research Laboratory, Kashiwa, Chiba 277-8582, Japan\\
$^{10}$Kavli IPMU (WPI), UTIAS, The University of Tokyo, Kashiwa, Chiba 277-8583, Japan
}

\pubyear{\the\year{}}

\begin{document}
\label{firstpage}
\pagerange{\pageref{firstpage}--\pageref{lastpage}}
\maketitle

\begin{abstract}
We explore how the fraction of red (quenched) galaxies varies in the dwarf galaxy regime (10$^{7}$ M$_{\odot}$ < $M_{\star}$ < 10$^{9.5}$ M$_{\odot}$), using a mass-complete sample of $\sim$5900 dwarfs at $z<0.15$, constructed using deep multi-wavelength data in the COSMOS field. The red fraction decreases steadily until $M_{\star}$ $\sim$ 10$^{8.5}$ M$_{\odot}$ and then increases again towards lower stellar masses. This `U' shape demonstrates that the traditional notion of `downsizing' (i.e. that progressively lower mass galaxies maintain star formation until later epochs) is incorrect -- downsizing does not continue uninterrupted into the dwarf regime. The U shape persists regardless of environment, indicating that it is driven by internal processes rather than external environment-driven mechanisms. 
Our results suggest that, at $M_{\star}$ $\lesssim$ 10$^8$ M$_{\odot}$, the quenching of star formation is dominated by supernova (SN) feedback and becomes more effective with decreasing stellar mass, as the potential well becomes shallower. At $M_{\star}$ $\gtrsim$ 10$^9$ M$_{\odot}$, the quenching is driven by a mix of SN feedback and AGN feedback (which becomes more effective with increasing stellar mass, as central black holes become more massive). The processes that quench star formation are least effective in the range 10$^{8}$ M$_{\odot}$ < $M_{\star}$ < 10$^{9}$ M$_{\odot}$, likely because the potential well is deep enough to weaken the impact of SN feedback, while the effect of AGN feedback is still insignificant. The cosmological simulations tested here do not match the details of how the red fraction varies as a function of stellar mass -- we propose that the red fraction vs stellar mass relation (particularly in the dwarf regime) is a powerful calibrator for the processes that regulate star formation in galaxy formation models.
\end{abstract}


\begin{keywords}
galaxies: formation -- galaxies: evolution -- galaxies: dwarf -- galaxies: star formation
\end{keywords}


\section{Introduction}

`Downsizing' is a fundamental concept in the existing galaxy evolution literature and refers to the observation that, in the mass regime $M_{\star}$ $\gtrsim$ 10$^{9.5}$ M$_{\odot}$, more massive galaxies appear to have completed their stellar mass buildup at earlier epochs than their lower mass counterparts. \citet{Cowie1996} were the first to articulate this effect by demonstrating that the maximum $K$-band luminosity (and hence stellar mass) of actively star-forming galaxies decreases smoothly with decreasing redshift. This implies that massive galaxies form the bulk of their stars at earlier cosmic epochs, while lower-mass systems continue to form stars more gradually over time. Subsequent studies have confirmed and expanded the downsizing picture using an array of different diagnostics. For example, \citet{Thomas2005} demonstrate that the most massive early-type galaxies exhibit high [$\alpha$/Fe] abundance ratios, consistent with rapid ($<1$ Gyr) formation timescales, while their lower-mass counterparts form stars over more extended periods of time. In a similar vein, \citet{Bundy2006} and \citet{Gu2018} show that the fraction of red (quenched) galaxies increases with stellar mass, and that the characteristic stellar mass above which most galaxies are quenched decreases with time. \citet{Pacifici2016} add support to this picture by reconstructing the star formation histories of galaxies via their spectral energy distributions and demonstrating that the average formation redshift decreases with decreasing stellar mass.

It is important to note, however, that the past downsizing literature is largely restricted to the high mass ($M_{\star}$ $\gtrsim$ 10$^{9.5}$ M$_{\odot}$) regime because these are the objects that are bright enough to be readily observable in past large surveys across cosmic time. In the high-mass regime the observational consensus is that more massive objects stop forming stars earlier, while progressively lower-mass systems quench at later epochs. Interestingly, at first sight, the pattern of star formation suggested by downsizing appears to be in conflict with the expectations from hierarchical structure formation in the currently accepted $\Lambda$CDM cosmology. In hierarchical galaxy formation models, smaller dark-matter halos (which host lower mass galaxies) collapse at earlier epochs and merge under the influence of gravity to form more massive ones \citep[e.g.][]{White1978,Blumenthal1984}. This is seemingly at odds with the observed (anti-hierarchical) downsizing trend of earlier star formation in more massive systems and later star formation in their less massive counterparts.

However, recent work has shown that, at least in the high mass regime, incorporating feedback, particularly from active galactic nuclei (AGN), can reconcile this apparent contradiction. For example, AGN feedback implemented in semi-analytical models \citep[e.g.][]{Bower2006,Croton2006} suppresses cooling flows in massive halos, effectively halting star formation and producing old, red galaxies early in cosmic history. This quenching mechanism naturally leads to a scenario where more massive galaxies stop forming stars earlier than their less massive counterparts. Cosmological hydrodynamical simulations, such as Illustris \citep{Vogelsberger2014}, EAGLE \citep{Schaye2015}, and Horizon-AGN \citep{Dubois2016,Kaviraj2017,Beckmann2017}, have implemented feedback and galaxy evolution in a fully cosmological framework, broadly reproducing the observed signatures of downsizing in the high mass regime.  

While the presence of a downsizing trend has been convincingly demonstrated in this high mass regime (and can be explained in theoretical models via the implementation of AGN feedback), the picture is less clear for low mass (dwarf) galaxies ($M_{\star}$ $\lesssim$ 10$^{9.5}$ M$_{\odot}$). A potential obstacle against probing whether downsizing continues into the dwarf regime is the fact that past large-scale surveys like the SDSS \citep{Alam2015}, which have underpinned many of the recent advances in our understanding of galaxy evolution, are relatively shallow. As discussed in detail in \citet{Kaviraj2025}, typical dwarfs are not bright enough to be detectable outside the local neighbourhood ($z>0.02$) in such shallow surveys. The dwarfs that do appear in these datasets typically host significant levels of star formation. These young stellar populations are needed to boost the luminosities of the dwarfs above the detection thresholds of shallow surveys like the SDSS, making them detectable in these datasets. 

However, this bias also makes these dwarf samples dominated by blue star-forming systems. This effect becomes more pronounced at progressively lower stellar masses, as a larger fraction of red (i.e. quenched) galaxies preferentially move out of the selection (see Figure 4 in \citealt{Kaviraj2025} and the related discussion in their Section 3). Not unexpectedly, using deeper surveys than the SDSS results in much larger overall red fractions \citep[e.g.][]{Tanoglidis2021,Thuruthipilly2023,Kaviraj2025}. Interestingly, even dwarfs that lie at significant distances from nodes, filaments and massive galaxies (i.e. are spatially very isolated) show substantial red fractions, indicating that much of the star formation in dwarfs may be regulated via internal processes like baryonic feedback and that residing in dense environments is not a prerequisite for quenching to take place (\citealt{Kaviraj2025}, see also \citealt{Bidaran2025}). 

In the context of probing the presence of downsizing this is particularly problematic, since we are unable to reliably measure the fraction of galaxies which are red (and therefore quenched). Indeed, a galaxy sample dominated by blue galaxies -- as is the case for dwarfs in shallow surveys outside the local neighbourhood -- may show a spurious downsizing trend because red galaxies are essentially missing from the population. Exploring downsizing in the dwarf regime outside the local neighbourhood therefore requires surveys that are both deep and wide which can be used to construct large statistical samples of dwarfs outside the very local Universe. While this will be routinely possible using surveys like the forthcoming Legacy Survey of Space and Time \citep[LSST, e.g.][]{Ivezic2019}, \textit{Euclid} \citep{Gardner2006} and those using the Roman Space Telescope \citep{Spergel2015} some deep data already exists in the 2 deg$^2$ COSMOS field which can provide insights into the downsizing issue in the dwarf regime. As we describe in Section \ref{sec:C2020_sample} below, the depth of the available data enables us to construct mass-complete samples of galaxies down to $M_{\star}$ $\sim$ 10$^{7}$ M$_{\odot}$), out to at least $z\sim0.15$. The shape of the galaxy stellar mass function \citep[e.g.][]{Wright2017} ensures that there are significant numbers of dwarf galaxies, even in a field with a modest area like COSMOS.

Finally, it is worth noting that properties like the red fraction are inextricably linked to the processes that regulate star formation in galaxies, which remain poorly explored, particularly in the dwarf regime. As such, a study of how these properties behave as a function of stellar mass offers valuable constraints on the implementation of these processes in galaxy evolution models. Indeed, the predicted properties of dwarfs in theoretical models are very sensitive to the details of the physical assumptions used in the modelling, making such empirical constraints fundamental to testing and improving our theoretical infrastructure \citep{Martin2025}. 

This paper is organized as follows. In Section \ref{sec:data}, we describe the observational and theoretical datasets that underpin this study. In Section \ref{sec:environment_metrics}, we define several metrics that we use to probe local environment. In Section \ref{sec:red_fraction}, we study the galaxy red fraction in the dwarf regime, as a function of stellar mass, environment and morphology. We compare our results to the existing observational literature and the predictions of cosmological simulations and discuss the implications of our findings in terms of the processes that dominate the quenching of star formation in different mass regimes. We summarise our findings in Section \ref{sec:summary}.


\section{Data}
\label{sec:data}

In this study we combine (1) physical parameters (stellar masses, photometric redshifts and rest-frame colours) from the COSMOS2020 catalogue \citep{Weaver2022} and the SDSS NASA Sloan Atlas \citep{Blanton2011}, (2) deep images from the Hyper Suprime-Cam (HSC), the Hubble Space Telescope (HST) and the James Webb Space Telescope (JWST) and (3) two high resolution cosmological hydrodynamical simulations. In the following sections we describe the properties of these datasets.


\subsection{A mass-complete galaxy sample from the COSMOS2020 catalogue}
\label{sec:C2020_sample}

The sample of galaxies that underpins this study is drawn from the Classic version of the COSMOS2020 catalogue. This provides physical parameters (e.g. stellar masses, photometric redshifts and rest-frame colours) for $\sim$1.7 million sources in the $\sim$2 deg$^2$ COSMOS \citep{Scoville2007} field, which is centered at 10h, +02$^{\circ}$. The physical parameters are calculated via the \textsc{LePhare} SED-fitting algorithm \citep{Arnouts2002,Ilbert2006} implemented on photometry from around 40 broad and medium band filters across the UV through to 4.5 $\mu$m from the following instruments: GALEX \citep{Zamojski2007}, MegaCam/CFHT \citep{Sawicki2019}, ACS/HST \citep{Leauthaud2007}, Hyper Suprime-Cam \citep{Aihara2019}, Subaru/Suprime-Cam \citep{Taniguchi2007,Taniguchi2015}, VIRCAM/VISTA \citep{McCracken2012} and IRAC/Spitzer \citep{Ashby2013,Steinhardt2014,Ashby2015,Ashby2018}. 

Object identification in this catalogue uses a detection image that incorporates optical ($i,z$) imaging from the Ultradeep layer of the HSC Subaru Strategic Program (HSC-SSP), which has a point-source depth of $\sim$28 mag \citep{Aihara2019}, around 10 mag fainter than the magnitude limit of the SDSS spectroscopic main galaxy sample \citep[e.g.][]{Alam2015}. Optical and infrared aperture photometry are extracted using the \textsc{SExtractor} \citep{Bertin1996} and \textsc{IRACLEAN} \citep{Hsieh2012} codes respectively. The availability of deep data across a wide wavelength baseline results in photometric redshift accuracies better than $\sim$1 and $\sim$4 per cent for bright ($i<22.5$ mag) and faint ($25<i<27$ mag) galaxies respectively. In our analysis below, we use measured stellar masses, photometric redshifts, rest-frame colours and star formation rates (SFRs) directly from the COSMOS2020 catalogue. 

Since we are specifically interested in measuring the fraction of red galaxies in our study, it is essential to have a mass-complete sample of galaxies. In other words, we must construct a galaxy sample that includes all objects, regardless of their colour and, therefore, star formation history. To construct a mass-complete sample we follow the methodology of \citet{Kaviraj2025}, who estimate the minimum redshift out to which a galaxy population of a given stellar mass is complete in COSMOS2020. This redshift is defined as that at which a purely-old `simple stellar population' (SSP) of a given stellar mass, that forms in an instantaneous burst at $z=2$, is detectable, at the depth of the HSC Ultradeep imaging in COSMOS (which underpins object detection in the COSMOS2020 catalogue). \citet{Kaviraj2025} use this purely-old SSP as a faintest `limiting' case, because real galaxies, which do not contain only old stars, will be more luminous than this limiting value. Thus, if this purely-old SSP is detectable at the depth of a given survey, then the entire galaxy population at a given stellar mass should also be detectable in the survey in question. The top panel of Figure 4 in \citet{Kaviraj2025} shows that the galaxy population in COSMOS2020 is likely to be complete down to stellar masses of 10$^{7}$ M$_{\odot}$, out to at least $z\sim0.15$. The objects that underpin this study are therefore galaxies with stellar masses greater than 10$^{7}$ M$_{\odot}$ which have redshifts less than $z=0.15$. 


\subsection{Morphological classification via visual inspection using JWST, HSC and HST images}
\label{sec:morphological_classification}

For every galaxy, we use visual inspection of three images -- a JWST F277W image, an HSC $gri$ colour composite image and an HST F814W ($i$-band) image. The HST images are taken from the COSMOS cutout service\footnote{https://irsa.ipac.caltech.edu/data/COSMOS/index\_cutouts.html}, the HSC images are created using the Python function \texttt{make\_lupton\_rgb} \citep[described in][]{Lupton2004} from the Python library \texttt{astropy} applied to the third data release of the HSC-SSP and the JWST cutouts are created from the reduction of JWST data described in \citet{Adams2024} and \citet{Conselice2025}. The angular resolutions of the JWST, HSC and HST images are 0.03, 0.6 and 0.03 arcseconds respectively. 

As noted in \citet{Lazar2024a}, given their intrinsic faintness, the features in dwarf galaxies may exhibit lower contrast than in their massive counterparts. Therefore, for each galaxy, we also create `unsharp masked' versions of the JWST, HST and HSC image. Unsharp masking \citep[e.g.][]{Malin1977} creates a blurred image by convolving the original with a kernel, which is then subtracted from the original. The difference image can then be multiplied by a factor which results in sharpening the edges of structures  contained within it. Unsharp masking has previously been used in astronomy to detect faint, low-contrast features like shells and tidal features inside and around nearby massive galaxies \citep[e.g.][]{Malin1983}. 

For the visual inspection, the images are randomised, both the original and unsharp-masked images of each galaxy are classified at the same time and physical parameters such as the stellar mass and redshift are kept hidden during the classification process to avoid introducing any biases. The visual inspection, carried out by one expert classifier (SK), separates the galaxy population into three broad morphological classes: early-type galaxies (ETGs), which have central light concentrations but otherwise smooth light distributions, late-type galaxies (LTGs) which lack a central light concentration that is typical of ETGs and often show structure (e.g. spiral arms, clumps etc.) and compact objects which, while resolved, are somewhat too small to classify securely. We note, however, that the shapes of these objects suggest that they are likely to be mostly ETGs. 

\begin{figure*}
\includegraphics[width=0.88\textwidth]{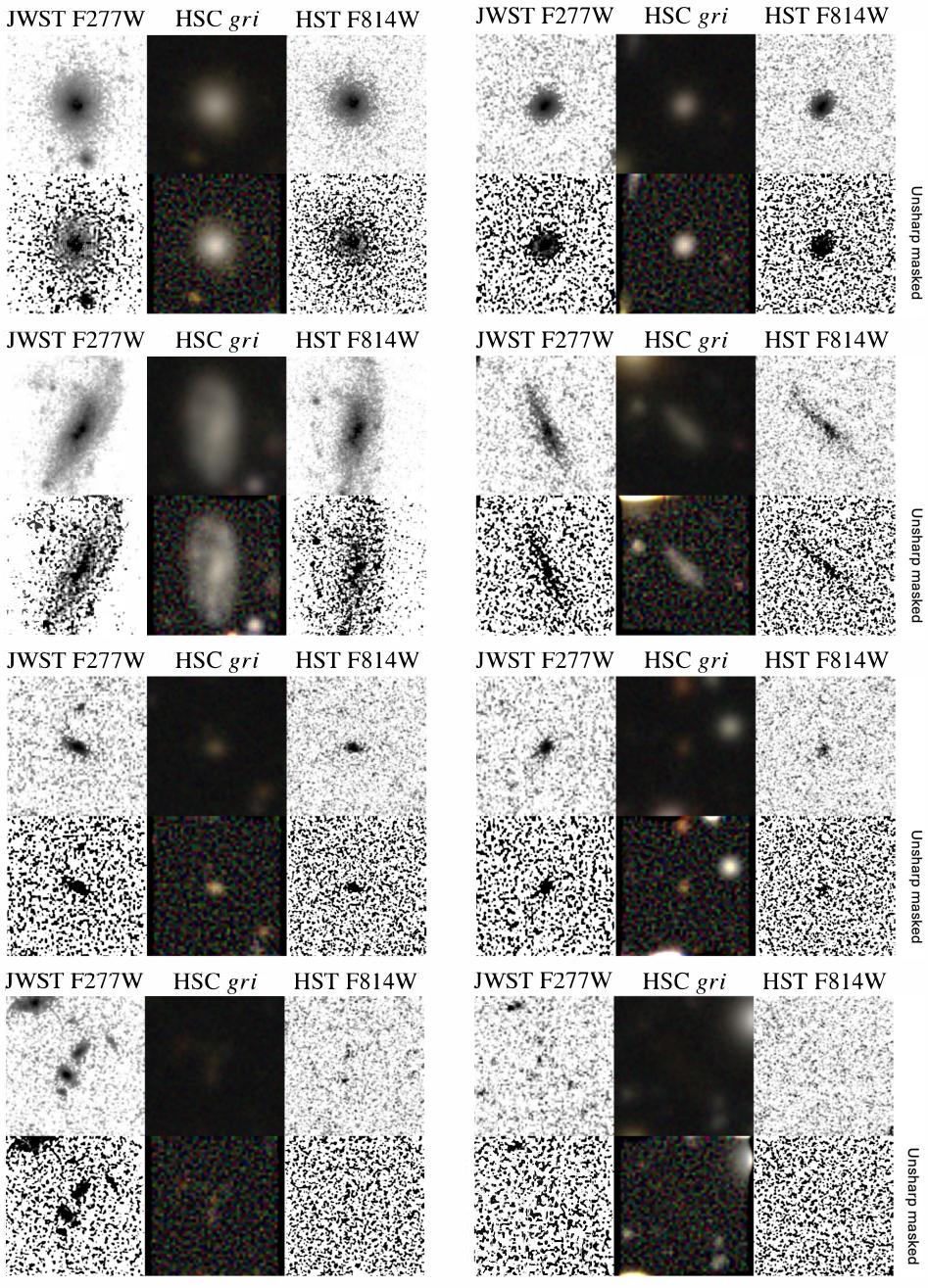}
\vspace*{2mm}
\caption{Example images of ETGs (row 1), LTGs (row 2), galaxies classified as compact (row 3) and those flagged as unclassifiable (row 4). The size of each image is 10 arcseconds on a side. The images from different instruments have different orientations - we intentionally keep it this way so that each galaxy is visually inspected at different orientations. In each image the first, second and third columns show the JWST, HSC and HST images respectively. Each individual image has six panels. Within each image the panels in the first row show the original images from the different instruments, while the second row shows their unsharp-masked counterparts.}
\label{fig:example_images}
\end{figure*}

A very small number of objects are labelled as unclassifiable. This is either because the objects do not have enough flux (particularly in the space-based images) to make classification possible or because they appear to be a blend of two objects in the ground-based HSC image, which would potentially compromise the COSMOS2020 photometry. Figure \ref{fig:example_images} presents examples of galaxies in our different morphological classes. 

Note that the JWST survey area covers 0.6 deg$^2$, around a quarter of the size of the COSMOS2020 footprint. Our morphological analysis is therefore restricted to this area. In Appendix \ref{app:jwst_subsample}, we demonstrate that the objects within the JWST footprint are an unbiased subset of the parent population by comparing the distributions of stellar mass, redshift, rest-frame $(g-i)$ colour and the star formation main sequence of the parent COSMOS2020 sample and the JWST subset. The distributions in all quantities are similar, indicating that the JWST galaxies are a relatively unbiased subset of the parent population. Our parent dwarf population contains $\sim$5900 galaxies, out of which $\sim$1000 reside within the JWST footprint (and therefore have morphological classifications). In the JWST subset around 97 per cent of objects are classifiable of which around 38, 57 and 2 per cent are classified as ETG, LTG and compact respectively. In what follows, the analysis presented in Section \ref{sec:red_fraction_morphology} is based on the JWST subset (since it relies on the morphological classifications described above), while all other analyses utilize the parent sample of dwarf galaxies.

\subsection{The NASA-Sloan Atlas}

The NASA Sloan Atlas (NSA)\footnote{https://nsatlas.org/} is a catalogue of the properties of nearby galaxies that combines data from the SDSS and GALEX \citep{Morrissey2007} surveys. The NSA provides a homogenized set of photometric and spectroscopic measurements for galaxies at redshifts $z < 0.055$ \citep{Blanton2011}, including UV imaging from GALEX, SDSS optical imaging and spectroscopy and various value-added parameters derived from these datasets. A key feature of the NSA is its reprocessing of SDSS images using improved background subtraction that may more accurately capture the properties of faint structures like dwarf galaxies and the outer regions of massive galaxies. Note that, while our key results are based on the COSMOS2020 dataset, we use the NSA to demonstrate the impact (particularly on incompleteness) of using a shallow survey. 


\subsection{The NewHorizon and TNG50 cosmological simulations}

We briefly describe the NewHorizon and TNG50 simulations whose predictions are compared to our observational results in Section \ref{sec:comparison_simulations}. Table \ref{tab:simulation_properties} summarizes the properties of each simulation. We provide an outline of their key characteristics below.

NewHorizon \citep{Dubois2021} is a high-resolution zoom-in of a spherical average-density region within the the parent Horizon-AGN simulation \citep{Dubois2014,Kaviraj2017}, which employs the adaptive mesh refinement code \textsc{RAMSES} \citep{Teyssier2002}. The simulated region has a diameter of 20 Mpc and achieves a maximum spatial refinement of 34 pc, with a stellar mass resolution of $1.3\times10^4$ M$_\odot$. Star formation is driven by dense, self-gravitating gas, with an efficiency regulated by local turbulence. The simulation implements mechanical supernova feedback from Type II supernovae \citep{Kimm2014} and the interstellar medium (ISM) is partially resolved, allowing for multi-phase structure to emerge. The parameters of the simulation are not explicitly calibrated to observational data, except for the efficiency of AGN feedback, which is chosen to reproduce the local $M_{\rm BH}-M_{*}$ relation. We direct readers to \citet{Dubois2021} for further details about the simulation. 

TNG50 \citep{Nelson2019a,Nelson2019b,Pillepich2019} simulates a 50 Mpc cosmological box using the moving mesh code \textsc{AREPO} \citep{Springel2010}. It employs magneto-hydrodynamics and is explicitly calibrated to match an array of key low-redshift observables, such as the galaxy stellar mass function and galaxy sizes in the intermediate and high-mass regime. Star formation is driven by a Kennicutt-Schmidt relation \citep[][]{Kennicutt1998} within an idealized two-phase ISM governed by an effective equation of state \citep{Springel2003}. The simulation implements feedback from AGN, Type I and Type II supernovae, and stellar winds through a combination of thermal and kinetic modes. Stellar feedback is implemented via kinetic winds and initially decoupled from the dense ISM. AGN feedback, on the other hand, includes both thermal energy injection in the high black-hole accretion rate regime and decoupled kinetic winds in the low-accretion regime. We direct readers to \citet{Nelson2019b} and \citet{Pillepich2019} for further details about the simulation.

It is worth exploring the types of environments that are likely to be present in the COSMOS footprint at $z<0.15$, by considering the \textit{M}$_{200}$ values (which are proxies for the virial masses) of groups identified in the literature \citep{Finoguenov2007,George2011,Gozaliasl2014,Gozaliasl2019}. At our redshifts of interest the virial masses of groups in COSMOS lie in the range 10$^{12.9}$ M$_{\odot}$ < \textit{M}$_{\rm{200}}$ < 10$^{13.8}$ M$_{\odot}$, with a median value of 10$^{13.1}$ M$_{\odot}$. For comparison, a small cluster like Fornax has a virial mass of $\sim$10$^{13.9}$ M$_{\odot}$ \citep{Drinkwater2001}, while large clusters like Virgo and Coma have virial masses of $\sim$10$^{15}$ M$_{\odot}$ \citep[e.g.][]{Fouque2001,Gavazzi2009}. Thus, the galaxy population considered in this study resides in relatively low-density environments (i.e. groups and the field) and not in rich clusters. The simulations we use here bracket the types of environments in the observations (and are also not focused on clusters), making the comparisons between data and theory consistent. 


\subsubsection{Calculating mock photometry for simulated galaxies}

In order to produce a consistent comparison between observations and theory, we follow the procedure described in \citet{Martin2022} to construct realistic mock HSC images and use them to calculate synthetic HSC photometry for galaxies in NewHorizon and TNG50. We then use the synthetic photometry in the HSC $g$ and $i$ bands to construct predicted red fractions which can be compared to our observational data. Here, we provide a brief outline of the key aspects of this methodology and refer readers to \citet{Martin2022} for further details. 

We start by creating a spectral energy distribution (SED) for each star particle within the galaxy in question, using a grid of simple stellar populations (which have a single age and metallicity) constructed using the \citet{Bruzual2003} population synthesis model. For each star particle, the grid is interpolated to its age and metallicity and its SED extracted assuming a \citet{Chabrier2003} initial mass function. We calculate the dust attenuation via a screen model which depends on the dust column density summed along the line of sight to each star particle, estimated using a gas-to-dust ratio of 0.4 \citep[e.g.][]{Draine2007}. A Milky Way dust grain model \citep{Weingartner2001} is then used to create the dust attenuated SED for the star particle. As demonstrated by \citet{Watkins2025} the low gas column densities of dwarf galaxies mean that changing the precise dust model has little impact on the broad-band photometry of galaxies in this mass regime. In 95 per cent of our simulated dwarfs, the $i$-band magnitudes are attenuated by less than 0.2 mag, with a majority being attenuated by a negligible amount.

We then convolve the SED of each star particle with the HSC transmission functions \citep{Kawanomoto2018}. This is followed by an adaptive smoothing of the galaxy star particles in 3D to better represent the distribution of stellar mass in phase space and remove unrealistic variations between adjacent pixels due to resolution effects. This smoothed distribution is then summed along one of the axes and binned into 0.168 arcsecond pixels to produce 2D $g$, $r$, and $i$-band flux maps. Finally, these are convolved with the HSC point spread function \citep{Montes2021} and AB magnitudes are calculated in each band of interest. We then use these magnitudes to construct rest-frame $(g-i)$ colours that can be compared to the observational data.

\begin{table*}
\caption{Key characteristics of the cosmological hydrodynamical simulations used in this study.}
\label{tab:simulation_properties}
\begin{tabular}{@{}lll@{}}
\toprule
                   & NewHorizon \citep{Dubois2021}                                                   & TNG50 \citep{Nelson2019b,Pillepich2019}                                                      \\ \midrule
Code               & RAMSES \citep{Teyssier2002}                                                                                          & AREPO  \citep{Springel2010}                                                                                                       \\
Volume             & Zoom of 20 Mpc spherical region                                                             & 50 Mpc box                                                                                                   \\
Spatial resolution & 34 pc                                                                                       & 100 -- 140 pc                                                                                              \\
Mass resolution    & $m_{\star}=1.3\times10^{4}$ M$_{\odot}$                                                            & $m_{\star}=8.5\times10^{4}$ M$_{\odot}$                                                                        \\
Environment        & Maximum halo mass $\sim 10^{13}$ M$_{\odot}$                                                                & Maximum halo mass $\sim 10^{14}$ M$_{\odot}$                                                                             \\
Star formation     & \begin{tabular}[c]{@{}l@{}}\hspace{-11pt} Dense, self-gravitating gas;\\ \hspace{-9pt}efficiency modulated by turbulence\end{tabular}                                                                           & \hspace{-10pt}\begin{tabular}[c]{@{}l@{}}Schmidt law in two-phase ISM;\\ density threshold-based\\ \citep{Springel2003}\end{tabular}                                                                                               \\
SN feedback        & \begin{tabular}[c]{@{}l@{}}\hspace{-10pt}Mechanical feedback from SN\\\hspace{-10pt}Type II \citep{Kimm2014}\end{tabular} & \begin{tabular}[c]{@{}l@{}}\hspace{-10pt}Direct heating and delayed kinetic winds\\\hspace{-10pt}from SNe I/II and stellar winds\\\hspace{-10pt}\citep{Springel2003}\end{tabular}  \\
ISM physics        & Partially resolved multiphase ISM                                                                & \begin{tabular}[c]{@{}l@{}}\hspace{-10pt}Idealised two-phase model with\\\hspace{-10pt}effective equation of state\end{tabular}                                                                               \\ \bottomrule
\end{tabular}
\end{table*}


\section{Metrics to measure (relative) environment}
\label{sec:environment_metrics}

We employ a series of metrics that trace the local environments of our galaxies in slightly different ways. Given that the galaxy redshifts available in our dataset are photometric in nature, we use a probabilistic approach which employs the overlap integral between the redshift probability distribution functions (PDFs) of both the galaxy in question and its neighbour. We describe each of our environment metrics below. In Section \ref{sec:red_fraction_environment}, we probe the effect of environment by splitting our galaxy population into different percentile ranges in each metric to explore how the red fraction varies as a function of changing environment. 


\subsection{Simple number density}

The simple number density calculates the projected number density of neighbours within an aperture centered at each galaxy's location and is described as follows:

\begin{equation}
    \scalebox{1.5}{$\rm\textit{$\Sigma$}_i = \displaystyle \textit{A}^{-1} \sum_j  \int \textit{P}_i(\textit{z}) \: \textit{P}_j(\textit{z}) \: d\textit{z} $},
	\label{eq:number_density_eq}
\end{equation}
where $\Sigma_{\rm i}$ is the measured number density of galaxy $i$ (i.e. the central galaxy), $A$ is the area of the aperture, $P_{\rm i}(z)$ is the redshift PDF of the central galaxy and $P_{\rm j}(z)$ is the redshift PDF of each neighbour. This metric is similar to traditional metrics employed in the literature to define local density \citep[e.g.][]{Baldry2006,Malavasi2016}. 


\subsection{Mass-weighted number density}

The mass-weighted projected number density weights the simple number density described above (Eq. \ref{eq:number_density_eq}) by the stellar mass of each neighbour within the aperture and is described as follows:

\begin{equation}
    \scalebox{1.5}{$\rm\textit{$\Sigma$}_i^{\rm mass} = \displaystyle \textit{A}^{-1} \sum_j \textit{M}_j \int \textit{P}_i(\textit{z}) \: \textit{P}_j(\textit{z}) \: d\textit{z} $},
	\label{eq:mass_density_eq}
\end{equation}
where $\Sigma_{\rm i}^{\rm mass}$ is the measured mass-weighted number density of galaxy $i$ (i.e. the central galaxy), $M_{\rm j}$ is the stellar mass of the neighbour and the other variables have the same meaning as in Eq. \ref{eq:number_density_eq} above. This metric also traces the number density but gives more significance to more massive neighbours.


\subsection{Tidal stress}

Since the tidal force due to a neighbor scales as $M_n/D^3_n$, where $M_n$ and $D_n$ are the mass of and the distance to the neighbour respectively, we define a metric that measures the total tidal stress on a galaxy due to all neighbours in a specified aperture as follows: 

\begin{equation}
    \scalebox{1.5}{$\rm\textit{TS}_i = \displaystyle log \Big( \sum_j \frac{\textit{M}_j}{\textit{$\theta$}^3_{ij}} \int \textit{P}_i(\textit{z}) \: \textit{P}_j(\textit{z}) \: d\textit{z} \Big) $},
	\label{eq:perturbation_index_eq}
\end{equation}
where $\rm\textit{TS}_i$ is the measured tidal stress on galaxy $i$ (i.e. the central galaxy), $\theta_{\rm ij}$ is the angular distance between the central galaxy and each neighbour and the other variables have the same meaning as in Eq. \ref{eq:mass_density_eq} above. We use the angular rather than the physical distance because the photometric redshift induces a potentially large uncertainty in calculating the latter. The tidal stress responds strongly to very nearby neighbours due to its $\theta^{-3}$ dependence.


\section{The galaxy red fraction across the dwarf regime}
\label{sec:red_fraction} 

\begin{figure}
\center
\includegraphics[width=\columnwidth]{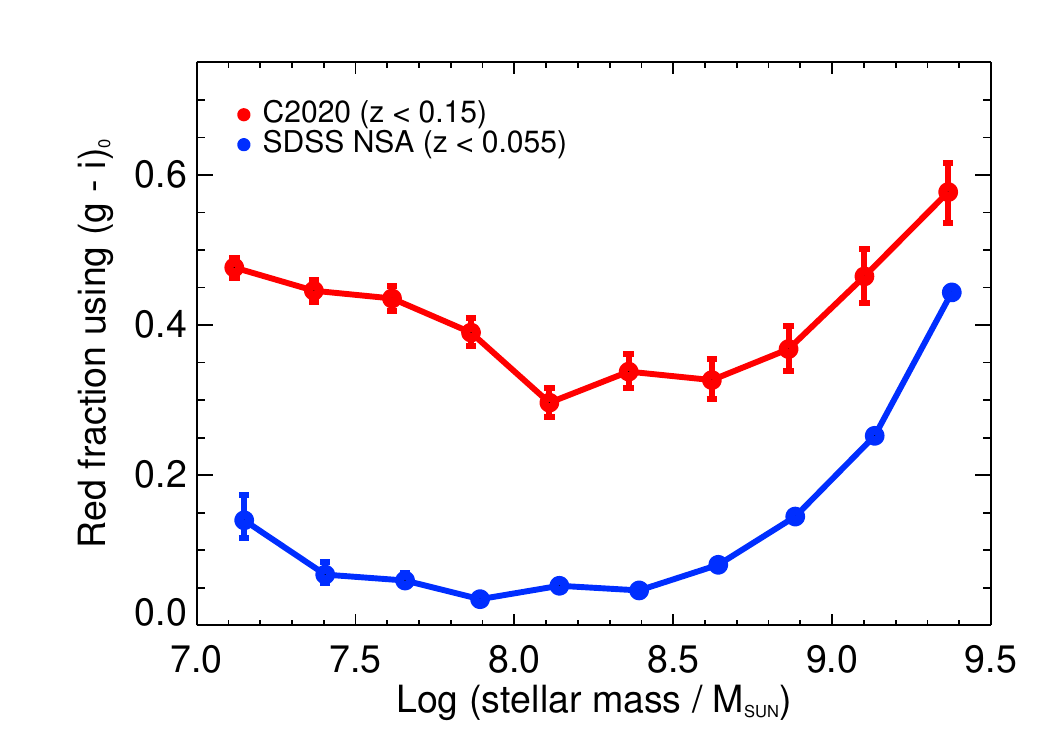}
\caption{The galaxy red fraction calculated using the rest-frame $(g-i)$ colour, as described in Section \ref{sec:red_fraction}, in COSMOS2020 (red) and the SDSS NSA (blue).} 
\label{fig:red_fraction}
\end{figure}

To calculate the red fraction in galaxies, we follow the methodology presented in \citet{Kaviraj2025} and \citet{Lazar2024a}, who use a threshold rest-frame $(g-i)$ value of 0.7 as a demarcation between red and blue galaxies in the dwarf regime. This is driven by the fact that the two dimensional galaxy distribution in colour -- stellar mass space separates into red and blue populations around this value in the dwarf population. Recall here that the population of red galaxies is thought to be driven of galaxies in which star formation has been quenched on timescales of around a Gyr in the past \citep[e.g.][]{Kaviraj2007}. Figure \ref{fig:red_fraction} presents the red fraction of galaxies in our COSMOS2020 sample (red curve) and its SDSS counterpart from the NSA (blue curve). The red fraction in COSMOS2020 exhibits a minimum around $M_{\star}$ $\sim$ 10$^{8.5}$ M$_{\odot}$ after which it rises towards lower stellar masses. In contrast the red fraction in the NSA evolves relatively monotonically (i.e. appears to show a spurious downsizing trend) and the value of the red fraction is several factors lower than that in COSMOS2020. Note that, at a lower redshift, the galaxy sample in any given survey will be less biased (and therefore more complete). Since the SDSS population in Figure \ref{fig:red_fraction} is restricted to a lower redshift than its COSMOS counterpart, the analysis favours the SDSS in terms of completeness. In spite of this, the SDSS red fractions are clearly underestimated. As discussed in the introduction, both of these differences are driven by the fact that the SDSS spectroscopic limit is too shallow to detect red dwarfs, a problem that becomes progressively worse at lower stellar masses.


In Appendix \ref{app:red_fraction_indicators}, we show how the measured red fraction changes if it is measured using the rest-frame ($u-z$) colour instead. We also show a `quenched fraction' that uses the quenching threshold defined in \citet{Kaviraj2025} -- this is a line that separates galaxies on the main locus of the star forming main sequence from those beneath it, which are then considered quenched (see Figure 6 in \citealt{Kaviraj2025}). It is clear that measuring the red fraction using different rest-frame colours gives virtually identical results and produces a result that is very similar to what is achieved using the quenched fraction, in which the quenched galaxies are identified in terms of their position in the SFR vs stellar mass plane. Since the quenched fraction depends on the SFR, which is, by definition, an intrinsic quantity, its correspondence with the red fraction indicates that the variation of the red fraction is driven by changes in the SFR and not by dust. In this context it is also worth noting that the low gas column densities in dwarf galaxies mean that dust reddening is expected to be negligible in this regime \citep[e.g][]{Martin2025,Watkins2025}.


The red fraction in COSMOS2020 demonstrates that galaxy downsizing does not continue uninterrupted into the dwarf regime. Around $M_{\star}$ $\sim$ 10$^{8.5}$ M$_{\odot}$, the red fraction turns over and starts increasing again towards lower stellar masses. In the following sections we study how the red fraction behaves as a function of environment and morphology and explore the origins of the observed trends.   

\begin{figure}
\center
\includegraphics[width=0.95\columnwidth]{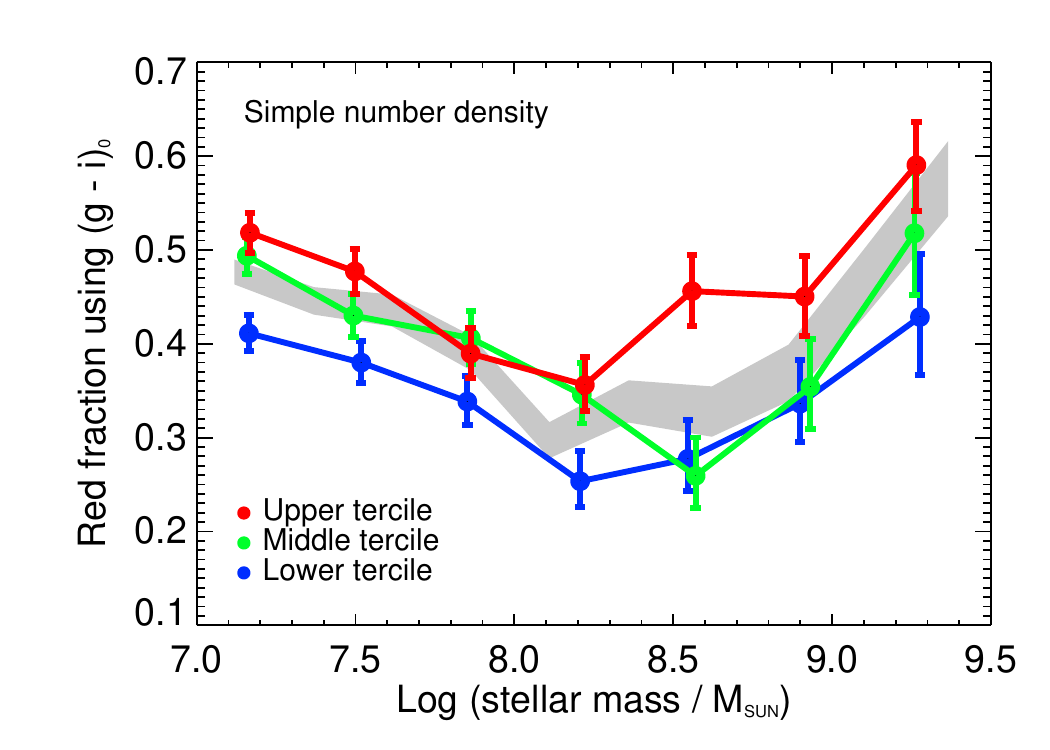}
\includegraphics[width=0.95\columnwidth]{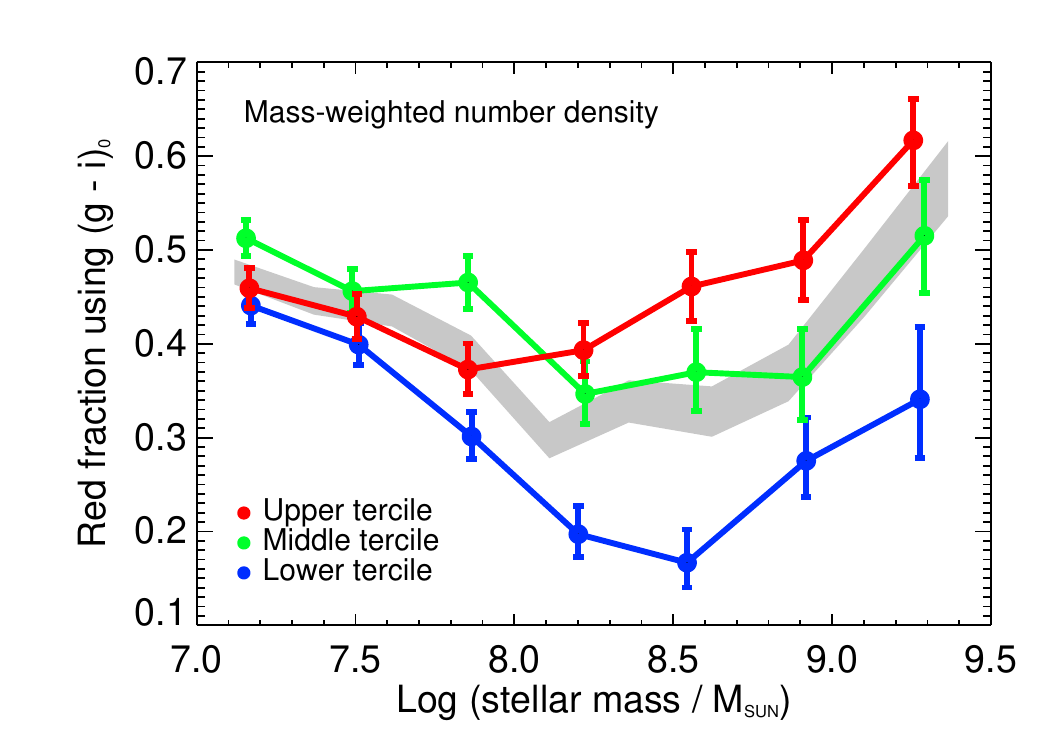}
\includegraphics[width=0.95\columnwidth]{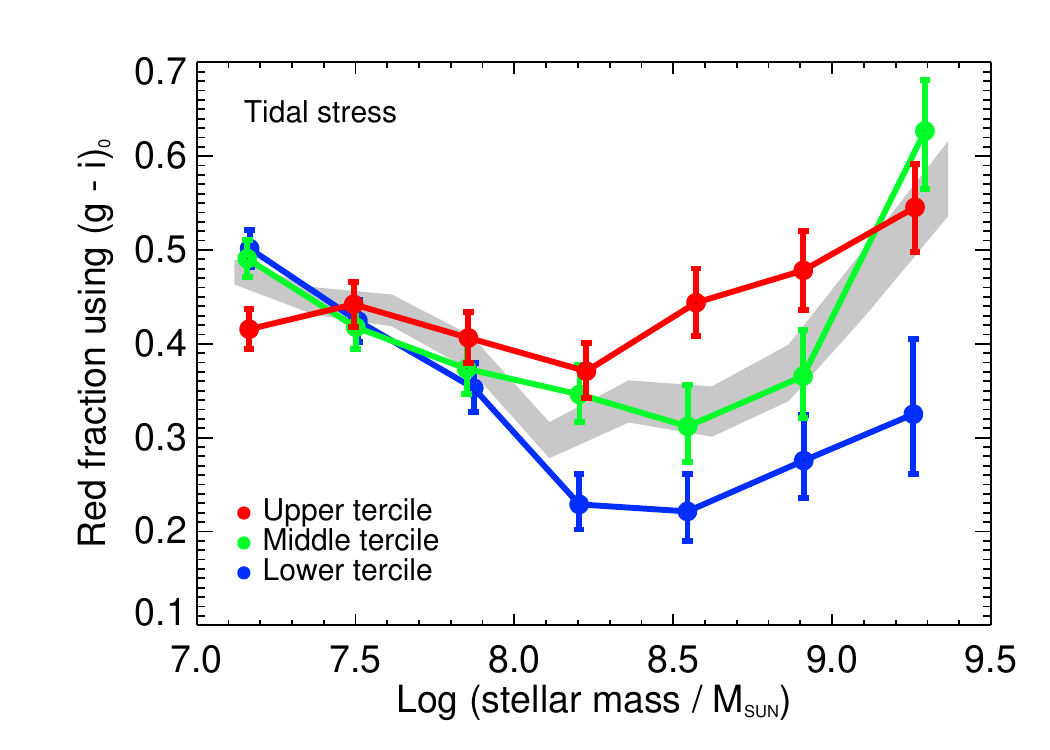}
\caption{The galaxy red fraction for different terciles in each of the environment metrics described in Section \ref{sec:environment_metrics}. The lower, middle and upper terciles contain galaxies in the 0 - 33rd, 33rd - 66th and 66th - 100th percentile values of the metrics in question. The top, middle and bottom panels show the red fraction as a function of the simple and mass-weighted number density and tidal stress respectively. The metrics are calculated within 0.5 Mpc radius apertures. The grey shaded region shows the red fraction of the parent sample of galaxies. Uncertainties are calculated following \citet{Cameron2011}.}  
\label{fig:red_fraction_terciles}
\end{figure}


\subsection{The red fraction as a function of environment}
\label{sec:red_fraction_environment}

We proceed by considering how the red fraction varies as a function of the three environment metrics described in Section \ref{sec:environment_metrics}. Recall that the simple number density measures the number density of neighbours within an aperture, the mass-weighted number density weights the aforementioned metric by the mass of the neighbour (and therefore favours high-mass neighbours) and the tidal stress responds strongly to very nearby neighbours. For each metric we split the dwarf galaxy population into terciles. The lower, middle and upper terciles contain galaxies in the 0th -- 33rd, 33rd -- 66th and 66th -- 100th percentile values of the metrics in question respectively. We then calculate the red fraction as a function of stellar mass separately in each tercile for every environment metric. It is important to note that, since the tercile boundaries are defined relative to our particular sample of galaxies, they represent \textit{relative} environmental categories.

Recent work demonstrates that large scale structure influences the properties of galaxies on spatial scales within around 0.5 Mpc \citep[e.g.][]{Mortlock2015,Malavasi2017,Etherington2017}. We therefore explore our environment metrics within apertures that trace angular separations of 0.5 Mpc. Figure \ref{fig:red_fraction_terciles} presents the red fraction as a function of stellar mass, split by the terciles in each metric, in apertures which have radii of 0.5 Mpc. Note that the results remain unchanged if larger apertures (e.g. 1 or 1.5 Mpc) are used.

When considering the simple number density (top panel in Figure \ref{fig:red_fraction_terciles}), we find that the U shape in the red fraction persists across all terciles of this metric, with galaxies in denser regions showing a roughly vertical positive offset (i.e. higher red fractions) from their counterparts in less dense regions. This is likely to be driven by the fact that processes like tidal interactions \citep{Moore1998,Martin2019,Jackson2021a} and ram pressure stripping \citep[e.g.][]{Balogh2000,Hester2006}, become more effective in denser regions, enhancing the removal of star-forming gas which reduces star formation activity even further and increases the probability of dwarfs becoming red\footnote{Note, however, that tidal perturbations can also induce transient bursts of star formation in dwarf galaxies \citep[e.g.][]{Martin2021}}. 

The behaviour is similar when the number density is weighted by the mass of the neighbours (middle panel of Figure \ref{fig:red_fraction_terciles}), except that the differences in the red fraction between the upper and middle terciles in this metric are much weaker. Nevertheless, all terciles exhibit a similar U shape in the red fraction, with the lower tercile showing lower overall red fractions than its middle and upper counterparts. Finally, for the tidal stress, the U-shape is present in the lower and middle terciles but is not as pronounced in the upper tercile. In addition, the value of the red fractions show less divergence at the lower end of the mass range considered here ($M_{\star}$ < 10$^{7.5}$ M$_{\odot}$). Recall, however, that the tidal stress is heavily skewed towards very nearby neighbours. 

The analysis above indicates that the U shape in the red fraction persists in virtually all terciles of the environment metrics considered in this study. In particular, it appears to persist in the lower tercile of each metric i.e. even in galaxies which inhabit the lowest density environments. 


\subsection{The red fraction as a function of morphology} 
\label{sec:red_fraction_morphology}

\begin{figure}

\center
\includegraphics[width=\columnwidth]{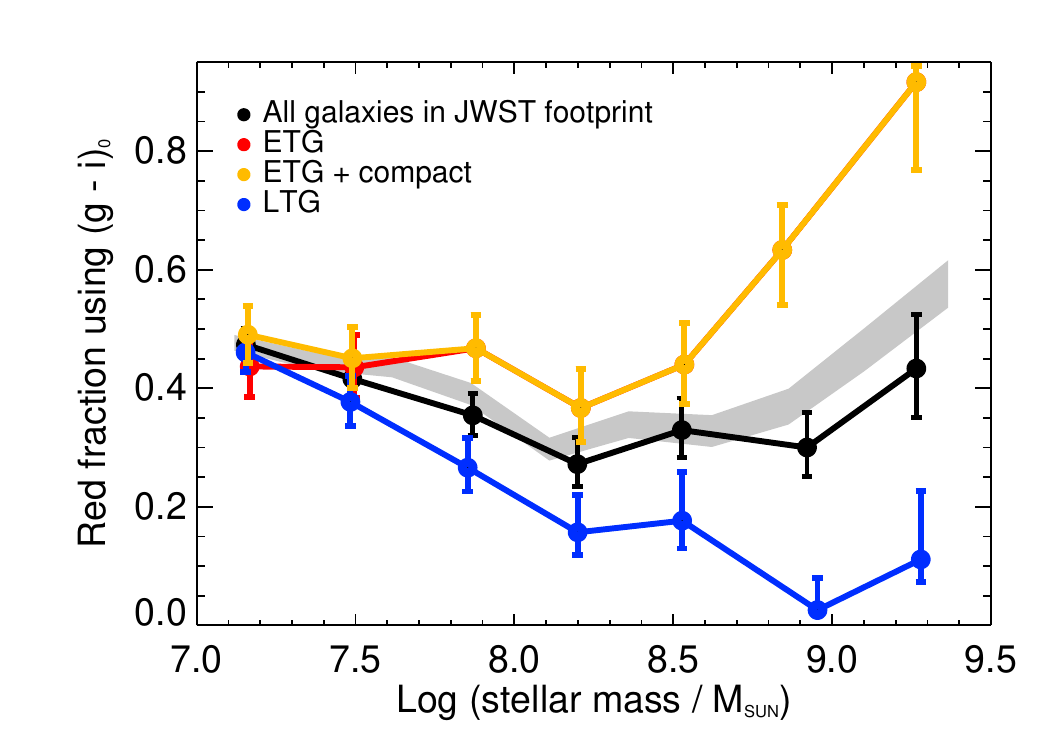}
\caption{The galaxy red fraction as a function of the different morphological classes described in Section \ref{sec:morphological_classification} - all galaxies in the JWST footprint (black) ETGs only (red), ETGs and compact object (orange) and LTGs (blue). The grey shaded region shows the red fraction of the parent sample of galaxies. Uncertainties are calculated following \citet{Cameron2011}.} 
\label{fig:red_fraction_morphology}
\end{figure}

Figure \ref{fig:red_fraction_morphology} shows how the red fraction behaves as a function of the three morphological types described in Section \ref{sec:morphological_classification} -- ETGs, LTGs and compact objects. Recall that the shape of the compact objects suggests that many of them are likely to be ETGs (although their small sizes makes it difficult to securely classify them). We show how the red fraction varies as a function of stellar mass for the full sample (grey region), all galaxies in the JWST footprint (black line), ETGs (red), LTGs (blue) and ETGs + compact galaxies (orange). Not unexpectedly, the compact objects are present (only) in galaxies with low stellar masses ($M_{\star}$ $\lesssim$ 10$^{7.7}$ M$_{\odot}$), where the red and orange curves diverge. Above this value the ETG + compact category is dominated by ETGs which results in the red and orange lines overlapping. 

We find that, while the U shape persists in all morphological classes, the position of the minimum appears to shift to larger stellar mass in the LTGs compared to the other morphological classes. We note that, while we would ideally also like to explore the behaviour of individual morphological classes in different environments, our sample size is too small for such an exercise (which therefore requires data from a larger survey like LSST).

\subsection{Comparison to the observational literature}

The non-monotonic behaviour of the galaxy red fraction and, in particular, an increase in the fraction of red/quenched objects with decreasing stellar mass in the dwarf regime has been reported in past work, although poor statistics and the general paucity of red objects in shallow surveys has hampered a definitive conclusion. For example, \citet{Slater2014} demonstrate a U shaped quenched fraction in the dwarf regime, both in the Local Group and in isolated field dwarfs, with the Local Group galaxies showing a roughly vertical positive offset towards higher red fraction values. This is similar to our findings, presented in Figure \ref{fig:red_fraction_terciles}. An increase in the quenched fraction with decreasing stellar mass has also been suggested in Local Group dwarfs by other studies \citep[e.g.][]{Weisz2015,Wetzel2015}, as well as in dwarfs that reside close to Milky Way like galaxies in the local neighbourhood \citep{Mao2021}, although the quenched fractions in the latter are lower than in their Local Group counterparts. A similar trend has also been reported in \citet{Karachentsev2013} in a catalogue of nearby galaxies (see the compilation in Figure 5 in \citealt{Weisz2015}). The position of the turnover (i.e. the bottom of the U shape) can appear in different locations in different studies. For example, while the turn over in \citet{Slater2014} and \citet{Karachentsev2013} occurs between 10$^8$ and 10$^9$ M$_{\odot}$, it occurs at larger stellar masses in \citet{Weisz2015}, possibly because the latter traces relatively higher density environments. It is worth noting here that the number count uncertainties in these small samples tend to be very large. Nevertheless, their results are in good qualitative agreement with our findings. 



\subsection{Comparison to cosmological simulations}
\label{sec:comparison_simulations}

\begin{figure}
\center
\includegraphics[width=\columnwidth]{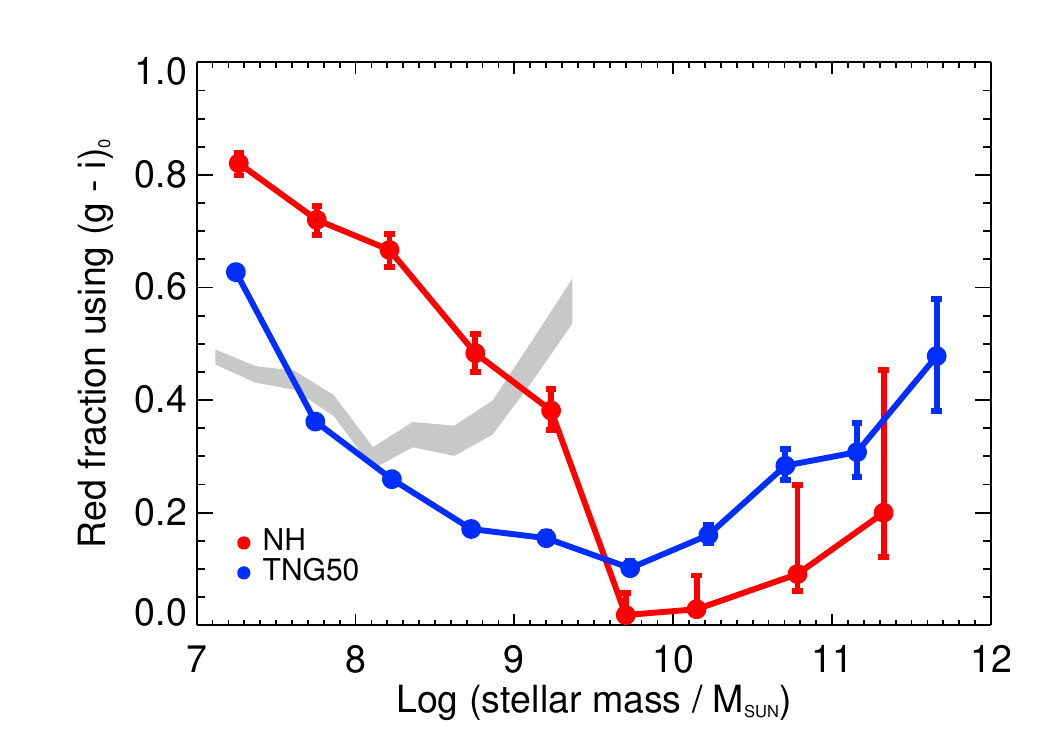}
\caption{Comparison of the red fraction in COSMOS2020 (grey shaded region) to predictions from the NewHorizon (red) and TNG50 (blue) cosmological simulations. Uncertainties are calculated following \citet{Cameron2011}.}
\label{fig:comparison_simulations}
\end{figure}

We now compare how the observed evolution of the red fraction with stellar mass compares with the predictions of the NewHorizon and TNG50 cosmological simulations. Figure \ref{fig:comparison_simulations} shows our observational results from COSMOS2020 and the corresponding curves from these two simulations. Note that the red fractions in the high mass regime from NewHorizon and TNG50 take into account the fact that the bimodality in the rest-frame $(g-i)$ colour shifts by around 0.1 mag because of the reddening caused by the increasing metallicity of galaxy populations towards higher stellar masses (see Figure 5 in \citealt{Kaviraj2025}).

It is apparent that the red fractions in both simulations diverge from the observations. The valus of the red fraction in NewHorizon are consistently too large across most of the dwarf regime. This suggests that the overall intensity of the processes that quench star formation (e.g. baryonic feedback) in NewHorizon may be too high in this regime. It is interesting to note here that the elevated red fraction in NewHorizon appears aligned with the fact that the sizes of simulated dwarfs in this simulation, particularly in the stellar mass range ($M_{\star}$ < 10$^{9}$ M$_{\odot}$), are somewhat too large compared to their observational counterparts \citep[e.g.][]{Martin2025,Watkins2025}. This discrepancy is likely to be driven by the fact that the bursty and variable star formation in NewHorizon produces stronger episodic outflows that preferentially eject central, low-angular-momentum gas \citep{Martin2025} and results in overquenching of star formation in shallower potential wells. 

However, NewHorizon galaxies also appear underquenched at $M_{\star}$ $\gtrsim$ 10$^{9}$ M$_{\odot}$. This is potentially driven by the fact that high density regions close to the galactic centre remain unresolved which underestimates the drag force. As a result, BHs tend to wander excessively from gas rich central regions, which reduces accretion from surroundings and diminishes the impact of AGN feedback (Han et al. in prep). There is a hint that the red fraction does eventually increase in the massive-galaxy regime but the small number of massive objects makes it difficult to draw precise conclusions.

In TNG50 the red fractions appear consistent with the observations across some of the dwarf regime (10$^{7.5}$ M$_{\odot}$ $\lesssim$ $M_{\star}$ $\lesssim$ 10$^{8.25}$ M$_{\odot}$) but, like in NewHorizon, galaxies appear underquenched outside this mass range in the dwarf regime. In particular, the quenching of star formation appears not to be efficient enough at the higher mass end of the dwarf galaxy population. The net result is that the minimum of the red fraction is inconsistent with the observations, as are the red fractions themselves across much of the dwarf regime. 

It is worth noting here that, in addition to the specific simulations that have been considered in detail here, modern high resolution simulations do produce U-shaped red/quenched fractions as a function of stellar mass \citep[e.g.][]{Oleary2023,Feldmann2023,Schaye2025}, when their predictions are considered across the dwarf and massive galaxy regimes. However, it is clear that the predictions of different simulations in the current literature can diverge, both with each other, and, more importantly, with their observational counterpart.

\subsection{The U-shaped red fraction - insights into the processes that regulate star formation in galaxies}
\label{sec:insights}

We complete our study by discussing the implications of our findings and the origin of the U-shaped galaxy red fraction. As a recap, our results demonstrate that downsizing does not continue uninterrupted into the dwarf regime, contrary to what would be expected from a simple extrapolation of what is known in massive galaxies into the dwarf regime. The relation between the red fraction and stellar mass shows a U shape, with a minimum around $M_{\star}$ $\sim$ 10$^{8.5}$ M$_{\odot}$ for the general dwarf population. The position of this minimum shifts to higher stellar mass ($M_{\star}$ $\sim$ 10$^{9}$ M$_{\odot}$) in dwarfs that have late-type morphology. The U shape in the red fraction appears to persist regardless of galaxy environment and, in particular, is present even in galaxies which reside in the lowest density environments across the different metrics explored in this study. Recall that the production of red galaxies is driven by the quenching of star formation. As a result, the red fraction is directly related to the mechanisms that regulate, and act to reduce, star formation activity in galaxies.   

The persistence of the U shape in all environments indicates that the processes that give rise to it do not depend on environment i.e. that they are internal in nature. However, while galaxies in all environments show a U shape, those in denser environments do tend to exhibit larger red fractions overall. This indicates that, while the shape of the curve (and therefore the process of star formation quenching) is determined primarily by internal mechanisms, denser environments clearly act to augment this process. This appears to be in line with some recent work which shows that many dwarfs in very isolated environments, e.g. at large distances from nodes, filaments and massive galaxies, are red \citep[e.g.][]{Kaviraj2025,Bidaran2025}.

The current consensus in the literature posits that most of the regulation of star formation in galaxies is driven by baryonic feedback from SN and/or AGN. The observed upturn in the red fraction below $M_{\star}$ $\sim$ 10$^{8}$ M$_{\odot}$ is likely to be driven by the fact that the impact of SN feedback becomes more effective in shallower potential wells, as it becomes easier for this process to heat and/or eject star-forming gas. By the same token, the upturn towards higher stellar masses, beyond $M_{\star}$ $\sim$ 10$^{9}$ M$_{\odot}$, is not consistent with this picture. However, it is worth noting in this context that recent work increasingly suggests that AGN feedback, which becomes more effective with increasing stellar mass \citep[e.g.][]{Kaviraj2007} likely makes a non-negligible contribution to star-formation quenching in dwarfs \citep[e.g.][]{Silk2017,Dashyan2018,Koudmani2019,Kaviraj2019,Davis2022,Koudmani2022}, in a similar fashion to how it is known to operate in massive galaxies \citep[e.g.][]{Beckmann2017}. It seems reasonable to suggest, therefore, that in the mass regime $M_{\star}$ > 10$^{8.5}$ M$_{\odot}$ a mix of SN and AGN feedback operates to regulate star formation, with the role of AGN feedback becoming proportionally more dominant at higher stellar masses.   

The observed differences between ETGs and LTGs appear aligned with this picture. First, LTGs are likely to have more inhomogeneous inter-stellar media which allow SN to create low-density channels through which gas can escape more easily. This enhances the efficiency of outflows and the impact of SN feedback in reducing star formation activity \citep[e.g.][]{Fragile2003,Ceverino2009,Martizzi2015} and may explain why the minimum of the red fraction curve shifts to larger stellar mass in the LTG population. Second, at a given stellar mass the red fraction in dwarf ETGs is typically larger than that in their LTG counterparts, with the difference increasing with increasing stellar mass. 

Past work has shown that, at least in the massive galaxy regime, the BH mass ($M_{\rm BH}$) in ETGs tend to be larger than in LTGs which have similar stellar masses \citep[e.g.][]{Davis2018}. In this regime, $M_{\rm BH}$ scales as $\sigma^4$, where $\sigma$ is the stellar velocity dispersion \citep[e.g.][]{Haring2004}. Since the galaxy stellar mass scales with $\sigma$ and the strength of AGN feedback is positively correlated with $M_{\rm BH}$ \citep[e.g.][]{Fabian2012}, the effectiveness of this process is likely to increase non-linearly with the galaxy stellar mass. Recent work has shown that the relationship between $M_{\rm BH}$ and stellar mass continues into the dwarf regime, possibly with a flatter slope \citep[e.g.][]{Martin-navarro2018,Ferre-mateu2021}, which suggests that the role of BH feedback is at least as strong or perhaps somewhat stronger in the dwarfs that form the basis of these studies. Overall, the increasing difference between the red fractions in dwarf ETGs and LTGs at larger stellar masses could be caused by a progressively larger contribution of AGN feedback to the quenching of star formation, consistent with the conclusions of recent empirical work that employs other techniques \citep[e.g.][]{Piotrowska2022,Bluck2023}. Note that a contribution to the larger red fractions observed in ETGs could also be made by `morphological quenching', where the presence of a bulge increases the gas velocity dispersion in the central regions, reducing gas fragmentation and consequently the SFR \citep[e.g.][]{Martig2009,Gensior2020}.

We note that strong reionization-driven suppression of star formation is unlikely to have significantly impacted the galaxy population in this study. The characteristic halo mass for such suppression is around 10$^{10}$ M$_{\odot}$ \citep[e.g.][]{Okamoto2008,Fitts2017} which translates to galaxy stellar masses around 10$^{7}$ -- 10$^{7.5}$ M$_{\odot}$. While this process may have some influence on the lower mass end of our sample, its impact is likely to be minor, given that the red fractions are significantly less than 1. 

In summary, our empirical results suggest that, in the dwarf population, stellar mass values between around 10$^8$ M$_{\odot}$ and 10$^{9}$ M$_{\odot}$ represent the regime where the processes that quench star formation become least effective (and star formation is therefore most efficient), because the potential well is deep enough to weaken the impact of SN feedback and the effect of AGN feedback is still minimal. Below these values the quenching of star formation becomes more effective with decreasing stellar mass as smaller potential wells find it harder to counteract the effect of SN feedback. Above these values a mix of SN and AGN feedback likely operates (augmented by processes like morphological quenching) which makes the quenching of star formation more effective as the galaxy mass (and consequently the size of the central BH) increases. While the U-shaped red fraction is not created by environmentally driven processes, denser environments do enhance the quenching of star formation and produce a similar U shape but shifted to larger red fractions.

In the context of the simulations considered here, our results suggest that the SN feedback prescriptions overquench dwarf galaxies in NewHorizon. In a similar vein, while the effect of SN feedback does produce realistic red fractions in the lower half of the dwarf mass regime in TNG50, the energy drawn from AGN feedback, i.e. through the growth of black holes, may not be sufficient in the upper half of the dwarf regime in this simulation.


\section{Summary}
\label{sec:summary}

We have explored how the galaxy red fraction evolves as a function of stellar mass in the dwarf galaxy regime, using a mass-complete sample of $\sim$5900 dwarf (10$^{7}$ M$_{\odot}$ < $M_{\star}$ < 10$^{9.5}$ M$_{\odot}$) galaxies which reside at $z<0.15$. The traditional view of downsizing posits that star formation persists until later epochs in lower mass galaxies. In past studies that probe the high mass regime ($M_{\star}$ > 10$^{9.5}$ M$_{\odot}$) downsizing indeed manifests itself via, for example, lower red and quenched fractions at progressively lower stellar masses and more extended star formation histories in lower mass galaxies. 

In the past, exploring whether downsizing exists in the dwarf ($M_{\star}$ $\lesssim$ 10$^{9.5}$ M$_{\odot}$) regime has been complicated by the fact that the wide area surveys that have dominated the past astrophysical literature (e.g. the SDSS) are relatively shallow. Typical dwarfs are too faint to be detectable in such surveys outside the very local Universe. The dwarfs that do appear in such shallow surveys (outside the local neighbourhood) are dominated by blue star-forming systems which, at a given stellar mass, are much brighter. However, using dwarf populations that are dominated by blue systems can produce a spurious downsizing effect that is driven by this selection bias. Probing the presence of downsizing in the dwarf regime therefore requires deep-wide surveys that can be used to create mass-complete samples of galaxies outside the very local Universe, as is the case in this study. Our main conclusions are as follows:

\begin{itemize}

    \item When a mass-complete sample of dwarfs is used, downsizing does not continue uninterrupted into the dwarf regime. For the general dwarf population, the galaxy red fraction decreases steadily until $M_{\star}$ $\sim$ 10$^{8.5}$ M$_{\odot}$ and then rises again towards lower stellar masses (creating a U shape). 
    
    \item The U shape persists regardless of (relative) environment, indicating that its origin is driven by internal processes rather than by driven by external environmentally-driven mechanisms. However, denser environments do augment the quenching process (even if they do not dominate it) by producing higher red fractions overall. 
    
    \item The U shape persists in all morphological types, although the minimum of the curve shifts to larger stellar masses ($M_{\star}$ $\sim$ 10$^9$ M$_{\odot}$) in late-type dwarfs.
    
    \item The processes that quench star formation appear to be least effective (and star formation is therefore most efficient) around stellar mass values between around 10$^8$ M$_{\odot}$ and 10$^9$ M$_{\odot}$. This is likely because the potential well is deep enough to weaken the impact of SN feedback, while the effect of AGN feedback is still insignificant. 
    
    \item Below 10$^8$ M$_{\odot}$ the quenching of star formation becomes more effective with decreasing stellar mass, as smaller potential wells find it harder to counteract the effect of SN feedback. Above 10$^9$ M$_{\odot}$ a mix of SN and AGN feedback likely operates (augmented by processes like morphological quenching) and the quenching of star formation becomes more effective as the galaxy mass (and therefore the BH mass) increases.

    \item The predicted red fractions as a function of stellar mass from the two cosmological hydrodynamical simulations tested here (NewHorizon and TNG50) do not converge, either with each other, or with their observational counterpart. 

    \item The photometric and structural properties of dwarf galaxies (drawn from unbiased, mass-complete samples) are valuable calibrators of the processes that regulate star formation in galaxy evolution models. Two scaling relations are particularly useful for this exercise. First, the variation of the red fraction as a function of stellar mass, as explored in this study, can provide constraints on the processes (e.g. feedback) that influence star formation activity in mock galaxies. Second, as discussed in \citet{Watkins2025}, the relationship between the size (or surface brightness) and stellar mass can put additional, complementary constraints on such processes, since they also influence the distribution of baryons in dwarf galaxies.

\end{itemize}

While our results are interesting in their own right, this study offers a preview into the enormous potential of new and future deep-wide surveys in enabling access to the dwarf regime and improving the state-of-the-art in our understanding of galaxy evolution. For example, photometric and imaging data from the LSST and \textit{Euclid} surveys would enable in principle, a similar study of hundreds of thousands of dwarf galaxies (compared to the few thousand employed here). The larger number of objects will alleviate cosmic variance, put our findings on a firmer statistical footing and are likely to lead to broader and more insightful conclusions about how the red fraction varies as a function of stellar mass, morphology environment and redshift. In future work, we aim to use these datasets to broaden the scope of the roadmap study that has been presented in this paper.


\section*{Acknowledgements}

We are grateful to the anonymous referee for many constructive comments that helped us improve the quality of the original manuscript. S. Kaviraj, IL and AEW acknowledge support from the STFC (grant numbers ST/Y001257/1 and ST/X001318/1). S. Kaviraj also acknowledges a Senior Research Fellowship from Worcester College Oxford. S. Koudmani is supported by a Research Fellowship from the Royal Commission for the Exhibition of 1851 and a Junior Research Fellowship from St Catharine's College Cambridge.

S.K.Y. acknowledges support from the Korean National Research Foundation (RS-2025-00514475; RS-2022-NR070872). This work was granted access to the HPC resources of CINES under the allocations c2016047637, A0020407637, and A0070402192 by Genci, KSC-2017-G2-0003, KSC-2020-CRE-0055, and KSC-2020-CRE-0280 by KISTI, and as a ``Grand Challenge'' project granted by GENCI on the AMD Rome extension of the Joliot Curie supercomputer at TGCC.

This research has made use of the Horizon cluster on which the simulation was post-processed, hosted by the Institut d'Astrophysique de Paris. We warmly thank S.~Rouberol for running it smoothly. This work is partially supported by the grant Segal ANR-19-CE31-0017 of the French Agence Nationale de la Recherche and by the National Science Foundation under Grant No. NSF PHY-1748958. 

The NASA-Sloan Atlas was created by Michael Blanton, with extensive help and testing from Eyal Kazin, Guangtun Zhu, Adrian Price-Whelan, John Moustakas, Demitri Muna, Renbin Yan and Benjamin Weaver. Renbin Yan provided the detailed spectroscopic measurements for each SDSS spectrum. David Schiminovich kindle provided the input GALEX images. We thank Nikhil Padmanabhan, David Hogg, Doug Finkbeiner and David Schlegel for their work on SDSS image infrastructure. Funding for the NASA-Sloan Atlas has been provided by the NASA Astrophysics Data Analysis Program (08-ADP08-0072) and the NSF (AST-1211644).

Funding for SDSS-III has been provided by the Alfred P. Sloan Foundation, the Participating Institutions, the National Science Foundation, and the U.S. Department of Energy. The SDSS-III web site is http://www.sdss3.org.

SDSS-III is managed by the Astrophysical Research Consortium for the Participating Institutions of the SDSS-III Collaboration including the University of Arizona, the Brazilian Participation Group, Brookhaven National Laboratory, University of Cambridge, University of Florida, the French Participation Group, the German Participation Group, the Instituto de Astrofisica de Canarias, the Michigan State/Notre Dame/JINA Participation Group, Johns Hopkins University, Lawrence Berkeley National Laboratory, Max Planck Institute for Astrophysics, New Mexico State University, New York University, Ohio State University, Pennsylvania State University, University of Portsmouth, Princeton University, the Spanish Participation Group, University of Tokyo, University of Utah, Vanderbilt University, University of Virginia, University of Washington, and Yale University.

The Galaxy Evolution Explorer (GALEX) is a NASA Small Explorer. The mission was developed in cooperation with the Centre National d'Etudes Spatiales of France and the Korean Ministry of Science and Technology.

The Hyper Suprime-Cam (HSC) collaboration includes the astronomical communities of Japan and Taiwan, and Princeton University. The HSC instrumentation and software were developed by the National Astronomical Observatory of Japan (NAOJ), the Kavli Institute for the Physics and Mathematics of the Universe (Kavli IPMU), the University of Tokyo, the High Energy Accelerator Research Organization (KEK), the Academia Sinica Institute for Astronomy and Astrophysics in Taiwan (ASIAA), and Princeton University. Funding was contributed by the FIRST program from the Japanese Cabinet Office, the Ministry of Education, Culture, Sports, Science and Technology (MEXT), the Japan Society for the Promotion of Science (JSPS), Japan Science and Technology Agency (JST), the Toray Science Foundation, NAOJ, Kavli IPMU, KEK, ASIAA, and Princeton University. This paper makes use of software developed for Vera C. Rubin Observatory. We thank the Rubin Observatory for making their code available as free software at http://pipelines.lsst.io/.

This paper is based on data collected at the Subaru Telescope and retrieved from the HSC data archive system, which is operated by the Subaru Telescope and Astronomy Data Center (ADC) at NAOJ. Data analysis was in part carried out with the cooperation of Center for Computational Astrophysics (CfCA), NAOJ. We are honored and grateful for the opportunity of observing the Universe from Maunakea, which has the cultural, historical and natural significance in Hawaii. This paper used data that is based on observations collected at the European Southern Observatory under
ESO programme ID 179.A-2005 and on data products produced by CALET and
the Cambridge Astronomy Survey Unit on behalf of the UltraVISTA consortium.


\section*{Data Availability}

The observational data used in this study are taken from \citet{Weaver2022}. 


\bibliographystyle{mnras}
\bibliography{references} 


\appendix

\section{Comparison of parent COSMOS2020 sample and the JWST observed subset}

\label{app:jwst_subsample}

\begin{figure*}
\center
\includegraphics[width=\columnwidth]{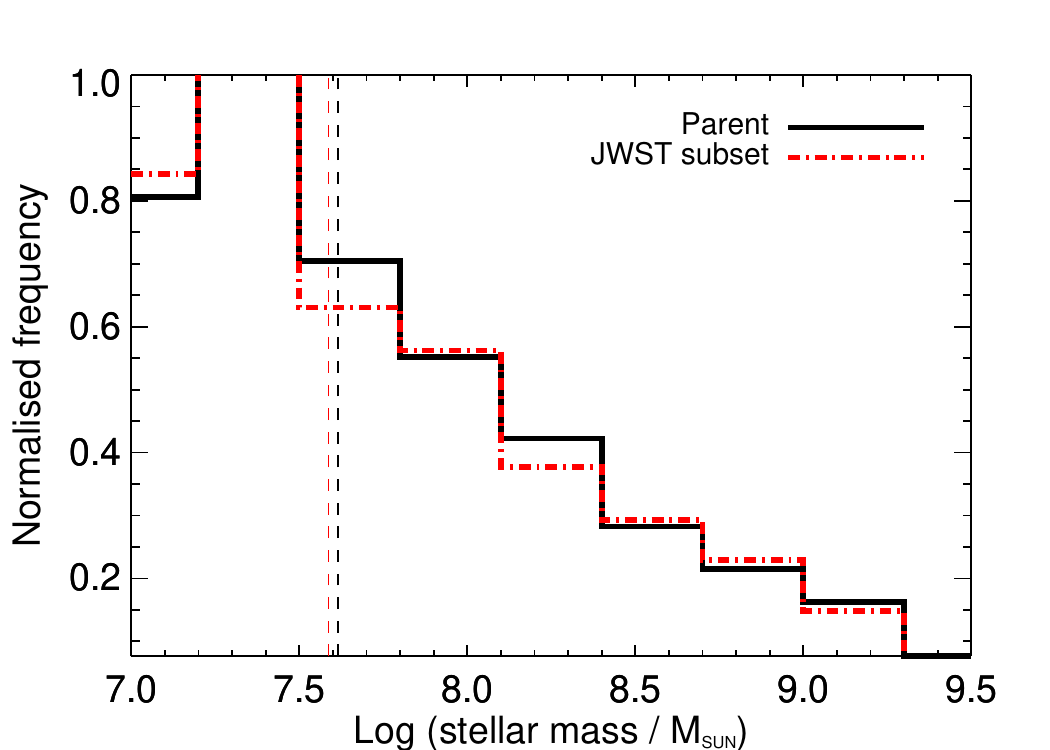}
\includegraphics[width=\columnwidth]{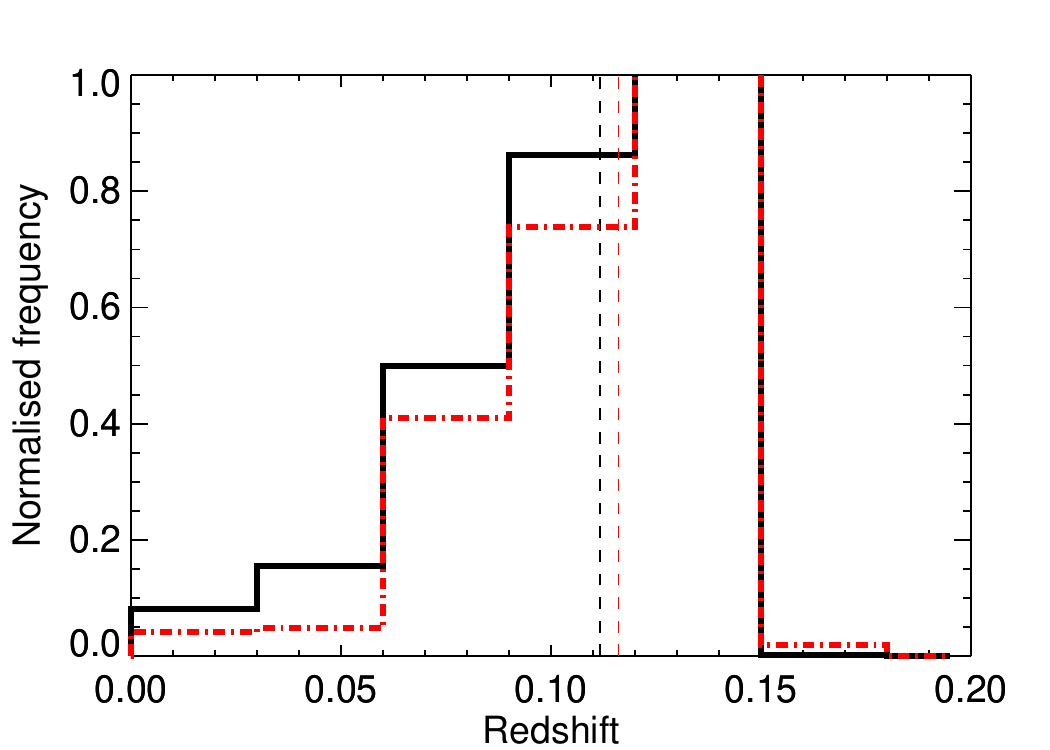}\\
\includegraphics[width=\columnwidth]{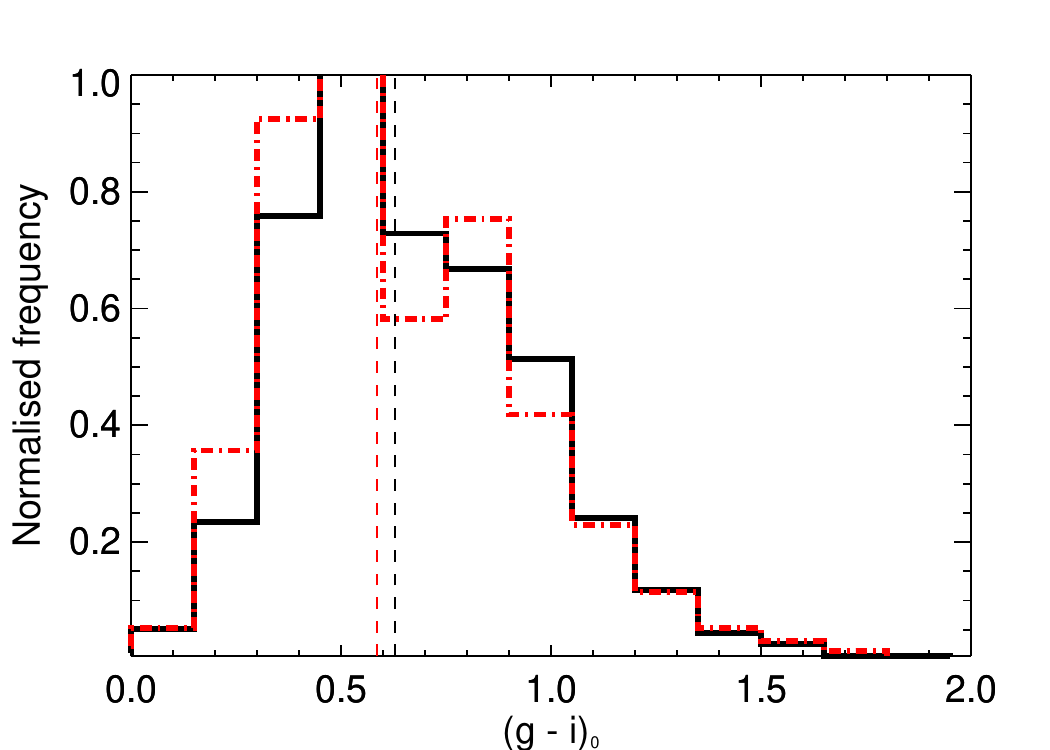}
\includegraphics[width=\columnwidth]{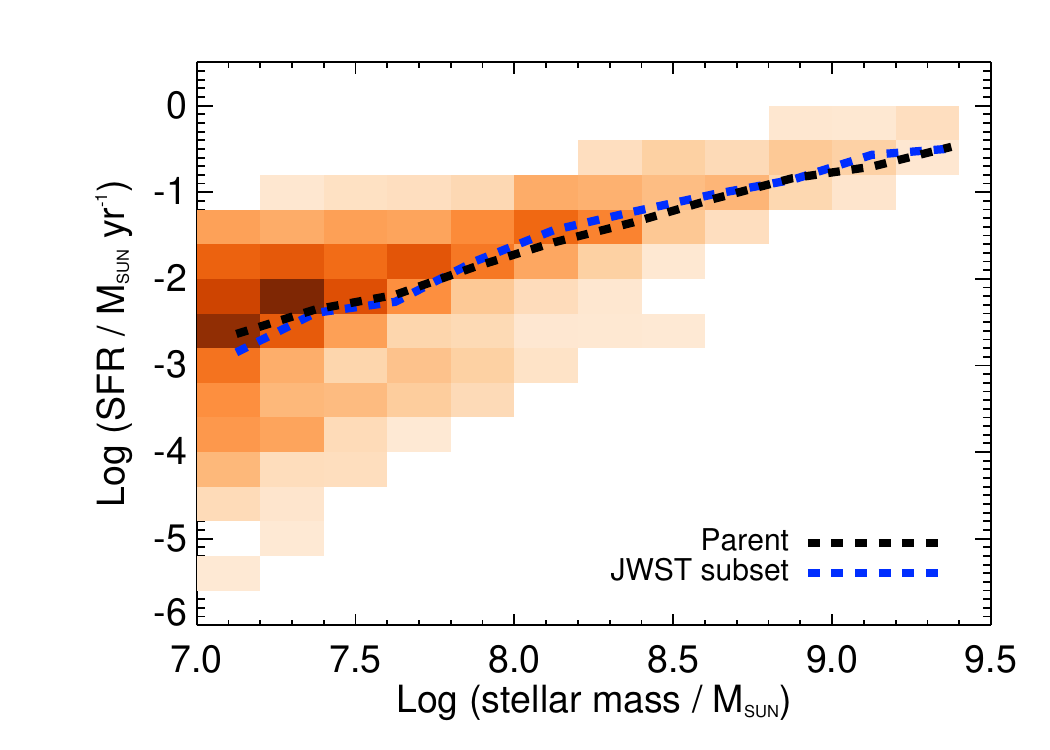}
\caption{Comparison of the distributions of stellar mass (top left), redshift (top right) and rest-frame $(g-i)$ colour (bottom left) in the full COSMOS2020 sample and the subset of galaxies in the JWST footprint. The vertical dashed lines show median values of the relevant distributions. The bottom right panel shows the star formation main sequence of the full sample as a heatmap and a running median of the SFR as a function of stellar mass for the full sample (black dashed line) and the JWST subset (blue dashed line). Darker colours in the heatmap indicate larger numbers of galaxies in a given region of this plane.} 
\label{fig:full_jwst_comparison}
\end{figure*}

In this section we compare the distributions of stellar mass, redshift, rest-frame $(g-i)$ colour and the star formation main sequence of the full COSMOS2020 sample and the galaxies that lie within the JWST footprint (Figure \ref{fig:full_jwst_comparison}). The properties of the COSMOS2020 population and the JWST subsample is similar in all quantities, indicating that the JWST galaxies are a random subset of the parent population.

\section{Measurement of the red/quenched fraction using different observables}

\label{app:red_fraction_indicators}

\begin{figure}
\center
\includegraphics[width=\columnwidth]{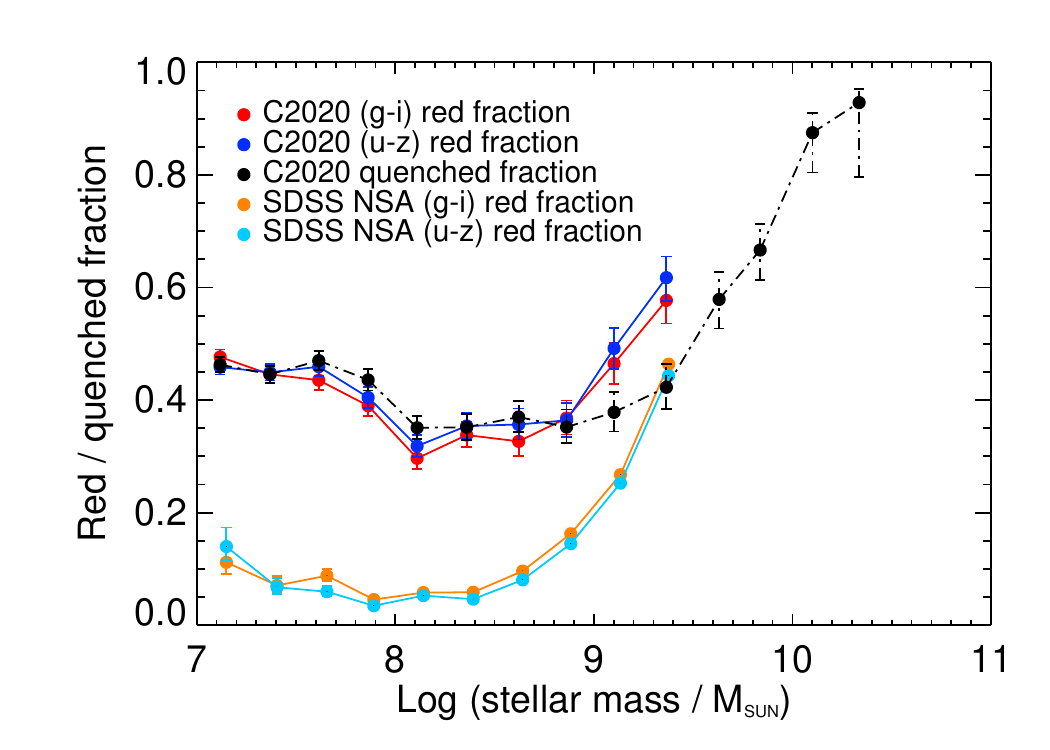}
\caption{Comparison of the red fraction calculated using the rest-frame $(g-i)$ colour in COSMOS2020 (red) and the SDSS NSA (dark blue) to the red fraction calculated using the rest-frame ($u-z$) colour in COSMOS2020 (orange) and the SDSS NSA (light blue). We also compare these red fractions to a `quenched fraction' (black), calculated using the quenching threshold described in \citet{Kaviraj2025}, which demarcates galaxies on the main locus of the star forming main sequence from those beneath it (which are then considered to be quenched). Uncertainties are calculated following \citet{Cameron2011}.} 
\label{fig:red_fraction_different_indicators}
\end{figure}

Figure \ref{fig:red_fraction_different_indicators} shows how the red fraction changes if it is measured using the rest-frame ($u-z$) colour. We also compare the red fraction to a `quenched fraction', calculated using the quenching threshold described in \citet{Kaviraj2025}, which demarcates galaxies on the main locus of the star forming main sequence from those beneath it (which are then considered quenched). Measuring the red fraction using different rest-frame colours gives virtually identical results and is similar to its quenched fraction counterpart.

\bsp
\label{lastpage}
\end{document}